# Asymmetry of collective excitations in electron and hole doped cuprate superconductors

W. S. Lee[1,*], J. J. Lee[2], E. A. Nowadnick[3,4], S. Gerber[1], W. Tabis[5,6], S. W. Huang[7], V. N. Strocov[7], E. M. Motoyama[3], G. Yu[5], B. Moritz[1], H. Y. Huang[8,9], R. P. Wang[8], Y. B. Huang[7,10], W. B. Wu[8], C. T. Chen[8], D. J. Huang[8], M. Greven[5], T. Schmitt[7], Z. X. Shen[1,2,3,*], and T. P. Devereaux[1,*]

[1]*SIMES, SLAC National Accelerator Laboratory, Menlo Park, CA 94025, USA*
[2]*Department of Applied Physics, Stanford University, Stanford, CA 94305, USA*
[3]*Department of Physics, Stanford, University, Stanford, CA 94305, USA*
[4]*Department of Physics, Columbia University, New York, NY 1002, USA*
[5]*School of Physics and Astronomy, University of Minnesota, Minneapolis, MN 55455, USA*
[6]*AGH University of Science and Technology, Faculty of Physics and Applied Computer Science, 30-059 Krakow, Poland*
[7]*Paul Scherrer Institut, Swiss Light Source, CH-5232 Villigen PSI, Switzerland*
[8]*National Synchrotron Radiation Research Center, Hsinchu 30076, Taiwan, R. O. C.*
[9]*Program for Science and Technology of Synchrotron Light Source, College of Science, National Tsing Hua University, Hsinchu 30076, Taiwan, R. O. C.*
[10]*Beijing National Laboratory for Condensed Matter Physics, and Institute of Physics, Chinese Academy of Sciences, Beijing 100190, China*

*To whom correspondence should be addressed: leews@stanford.edu, zxshen@stanford.edu and tpd@stanford.edu

**High-temperature superconductivity (HTSC) mysteriously emerges upon doping holes[1] or electrons[2] into insulating copper oxides with antiferromagnetic (AFM) order. It has been thought that the large energy scale of magnetic excitations, compared to phonon energies for example, lies at the heart of an electronically-driven superconducting phase with high transition temperatures ($T_c$)[3-5]. Comparison of high-energy magnetic excitations of hole- and electron-doped superconductors in connection with the respective $T_c$ provides an exceptional, yet un-capitalized opportunity to test this hypothesis[6-9]. Here, we use resonant inelastic x-ray scattering (RIXS) at the Cu $L_3$-edge[10,11] to reveal high-energy collective excitations in the archetypical electron-doped cuprate $Nd_{2-x}Ce_xCuO_4$ (NCCO)[2]. Surprisingly, despite the fact that the AFM correlations are short-ranged[12], magnetic excitations harden significantly across the AFM-HTSC phase boundary, in stark contrast with the hole-doped cuprates[6,7]. Furthermore, we find an unexpected and highly dispersive branch of collective modes in superconducting NCCO that are absent in hole-doped compounds. These modes emanate from zone center and weaken with increasing temperature, which signal a quantum phase distinct from superconductivity. The asymmetry uncovered between electron- and hole-doped cuprates provides new, unexpected dimensions to collective excitations that are generally important to the mechanism of superconductivity in these materials.**

Hole-doped cuprates display compelling evidence for the surprising persistence of magnetic excitations beyond the AFM phase boundary[6,7], as well as the existence of symmetry-broken phases, such as charge density waves[13,14,15] and orbital loop currents[16], distinct from superconductivity. Whether these are universal and exist on the other side of the cuprate phase diagram, i.e. with electron-doping, remains an important open question. To address this issue, collective excitations in antiferromagnetic ($x$ = 0.04) and superconducting ($x$ = 0.147 and 0.166) Nd$_{2-x}$Ce$_x$CuO$_4$ (NCCO) compounds are investigated via RIXS, uncovering unexpected behaviors.

We first discuss the types of low energy excitations seen in high-resolution RIXS in antiferromagnetic NCCO ($x$ = 0.04). As shown in Fig. 1a, a distinct feature emerges from the elastic peak (centered at zero energy transfer) at zone center (the Γ point). It disperses toward higher energy with increasing momentum (denoted as $q_{//}$, see Supplementary Information for the scattering geometry), and reaches maxima along high-symmetry directions at the AFM zone boundary (π, 0) and (π/2, π/2). The geometry of our experiment (σ-polarization-grazing-incidence branch, $q_{//}$ > 0; π-polarization-grazing-exit branch, $q_{//}$ < 0) allows us to identify this peak with single-magnon excitations, similar to those found in undoped AFM insulating compounds[17,18]. By fitting the spectra to Gaussians as depicted in Fig. 1b, the peak position of magnon excitations can be determined and their energy-momentum dispersion is extracted (Fig. 1c). It is well fit by linear spin waves of the two-dimensional nearest-neighbor Heisenberg model $\omega_m = \frac{\sqrt{2}}{a} c_s \sqrt{1 - \left(\frac{cosq_x + cosq_y}{2}\right)^2}$ (dashed line in Fig. 1c), where $a$ and $c_s$ are the lattice constant and spin-wave velocity, respectively. The band width of magnetic excitations (~ 300 meV) and the extracted value of $c_s$ = 830 ± 9 meV·Å are consistent with those found in other AFM cuprates[17,18,19].

Figure 2a presents an overview of the RIXS spectra in superconducting NCCO ($x$ = 0.147), where we again observe a dispersive feature (red markers) that reaches a maxima at the AFM zone boundary. Since AFM correlations become short-ranged at this doping level[12], these excitations are not magnons, but rather paramagnons with a short lifetime, as indicated by broad spectral widths which are comparable to their energies (See Supplementary Data). Unexpectedly, in addition to the paramagnon excitations, another excitation branch (blue ticks) is clearly resolved near the Γ point. As shown in the energy-momentum spectral intensity map with finer momentum steps (Fig. 2b), these excitations are strongest at the Γ point, where the paramagnon structure factor is zero. Moving away from the Γ point, the modes disperse to higher energies and weaken, becoming undetectable for $|q_{//}| > 0.3\pi$. The energy positions of the paramagnons and the additional excitation branch can be determined by fitting the spectra with Gaussians (Fig. 2c) to extract their energy-momentum dispersions, respectively (Fig. 2d).

One key discovery from this work is the finding that the paramagnon dispersion extends to much higher energies in the superconducting compound than in the AFM compound, as indicated in Fig. 2d. The AFM zone boundary energy of these paramagnons is approximately 450 meV for $x = 0.147$, a hardening of approximately 50% with respect to that of $x = 0.04$, also evident directly from the raw RIXS spectra (Fig. 3a). The hardening of the magnetic excitations was reported previously by Wilson et al.[20] using inelastic neutron scattering on the $Pr_{1-x}LaCe_xCuO_4$ family of electron-doped cuprates, where an increase of spin-wave velocity near (π, π) was found upon doping. However, this result was challenged by Fujita et al.[21], who did not observe any spin-wave-like dispersion up to 180 meV. Our result, which covers a wide range of the Brillouin zone away from (π, π), crucially confirms the hardening of magnetic excitations in the electron-doped cuprates. Notably, this behavior differs from that exhibited by hole-doped cuprates, for which the magnetic excitation spectrum slightly softens with doping[6,7]. Naïvely, one would expect the magnetic excitations to significantly soften upon either hole or electron doping away from the AFM parent compounds, since the doped charge carriers suppress long-range AFM order in both cases. In addition, since in-plane lattice constants change less than 0.5% from undoped to optimally-doped compounds[8], the pronounced magnetic excitation hardening cannot be attributed to structural modifications. We note that this surprising disparity with respect to electron- and hole-doping is inherent in the single-band Hubbard model, as shown by our determinant quantum Monte Carlo simulations (see Supplementary Theory) summarized in Fig. 3b.

The other important discovery of our work is the observation of collective modes emanating from zone center in the superconducting compounds (blue markers in Fig. 2d). The dispersion of these modes in the $x = 0.147$ compound can be fit with a standard form of charge collective modes $\omega_x = \sqrt{(E_\Gamma)^2 + (v_x \cdot q_{//})^2}$ (blue dashed curve in Fig. 2d) with the extracted zone center energy $E_\Gamma = 300 \pm 30$ meV, and velocity $v_x \sim 3.0$ eV·a/π, comparable to the Fermi velocity[22] $v_F \sim 2.0$ (eV·a/π). Although $v_x \sim v_F$ suggests that the modes may be of charge character, conclusive determination of whether they are charge or magnetic excitations requires a polarization analysis of the spectra[10], which is presently unavailable with comparable resolution. The existence of these collective modes is further confirmed by their observation in a slightly overdoped superconducting compound ($x = 0.166$, Fig. 4a). As demonstrated in Fig. 4b, the dispersion velocity is similar to that found in the $x = 0.147$ compound; however, $E_\Gamma$ of the $x = 0.166$ compound appears to shift toward a lower energy that cannot be resolved due to the limit of our instrument resolution. Nevertheless, our results unambiguously prove the existence of these collective modes, which are absent in heavily underdoped NCCO (i.e. $x = 0.04$ compounds shown in Fig. 1) and in any of the hole-doped superconducting cuprates[6,7,17].

What is the origin of these collective modes? The dispersion of these modes is reminiscent of that of plasmons; however, optical measurements[23] do not detect any plasmonic features at the energy scale comparable with $E_\Gamma$. Neither do these modes originate from bi-magnon nor higher-order magnetic excitations, since they are absent in the AFM $x = 0.04$ compound. Furthermore, it is unlikely that the new modes are directly associated with the superconducting state, since the superconducting gap[8,24] (~ 5 meV) is much smaller than $E_\Gamma$ and the modes persist well above the superconducting transition temperature (Fig. 4c).

On the other hand, the temperature and doping dependence of the collective modes near the $\Gamma$ point suggest that they might be associated with a symmetry-broken state other than superconductivity. As shown in Fig. 4c and d, collective modes near the $\Gamma$ point weaken with increasing temperature in a doping dependent manner. While the modes in the x = 0.166 compound vanish at temperatures higher than $T_\Gamma$ ~ 240 K, the zone center mode of the x = 0.147 compound persists even to room temperature, indicating that $T_\Gamma$ increases with underdoping. The doping-dependent $T_\Gamma$ cannot be explained by thermal broadening; rather, it is likely associated with the emergence of a doping-induced symmetry-broken state, which vanishes or may become fluctuating at high temperatures. Furthermore, the considerable decrease of $E_\Gamma$ with doping from $x = 0.147$ to $x = 0.166$ is also consistent with the mass of the collective mode changing near a quantum critical point (QCP) that is associated with this symmetry-broken state. Notably, $T_\Gamma$ appears to be higher than characteristic temperatures deduced from other measurements [8,25,26]; however, this is reminiscent of the existence of multiple temperature scales[14] and the well-known discrepancy of the "pseudogap" temperature $T^*$ in the literature for hole-doped cuprates, suggesting that a similarly complicated phenomenon is at play in electron-doped compounds. Whether the putative symmetry-broken state in NCCO is a charge/spin density wave[13,14,15], or a more exotic order, such as $d$-density wave[27] or intra-unit-cell orbital currents[16,28], remains an open question requiring further investigation.

Our results crucially complement existing knowledge, allowing us to depict a more complete picture regarding the doping evolution of the collective modes in doped cuprates on both sides of the phase diagram. As sketched in Fig. 5, upon hole-doping, the bandwidth of magnetic excitations (as determined by the AFM zone boundary energy) slightly softens, and a spin incommensurability plus a spin gap near ($\pi$, $\pi$) emerges at low energies[9]. This contrasts with the magnetic excitations observed in the optimally electron-doped superconducting compound[8,9,24], where the similarly determined magnetic bandwidth hardens significantly and no spin incommensurability develops near ($\pi$, $\pi$). In addition, the rapidly dispersive collective modes near zone center seen in electron-doped superconducting NCCO are absent in the hole-doped compounds. On the other hand, the implication of the existence of a quantum critical point beyond the AFM-SC phase boundary is reminiscent of the QCPs exhibited in the hole-doped counterparts[29,30], suggesting that a QCP located in proximity to the HTSC phase appears to be generic for both electron- and hole-doped cuprates.

Despite the common belief that magnetic excitations are crucial to HTSC[3,4,5], the significantly larger magnetic bandwidth in the superconducting electron-doped compounds does not correspond to a higher superconducting transition temperature ($T_c$). The spectral weight of the paramagnon remains significant in the superconducting compounds (see Fig. 3a and Supplementary Data), indicating that it is not the reason for the lower $T_c$. Thus, it is important to identify which factors, the magnetic excitations near ($\pi$, $\pi$), the underlying Fermi-surface topology, or additional effects, are not optimized in the electron-doped cuprates for the purpose of achieving a higher $T_c$. Finally, whether the observed collective modes near the $\Gamma$ point are beneficial or detrimental to the formation of superconducting pairs remains an important open question.

**Methods:**

Single crystals of $Nd_{2-x}Ce_x CuO_4$ were grown by traveling-solvent floating-zone (TSFZ) methods. Antiferromagnetic ($x = 0.04$), near optimally-doped superconducting crystals $x = 0.147$ and $x = 0.166$ were selected for our measurements. The superconducting transition temperature of $x = 0.147$ and $0.166$ compounds are 25 K. The doping levels of the superconducting samples were further characterized by EDS/SEM (see Supplementary Method). The lattice constants for the NCCO crystals are $a = b = 3.9$ Å, and $c = 12.1$ Å. The data shown in Fig. 1, Fig. 2, and Fig. 3 were taken at the ADRESS beamline, Swiss Light Source (SLS, Switzerland); Some of the data shown in Fig. 4 were taken at the beamline BL05A1, National Synchrotron Radiation Research Center (NSRRC, Taiwan) using the newly constructed AGM-AGS spectrometer[31]. All the data were taken with an energy resolution of approximately 130 meV, and with the scattering angle set at 130 degrees. The sample (001) surface was prepared by *in-situ* cleaving. The RIXS spectra shown in Fig. 1, Fig. 2, and Fig. 3 were normalized by the incident photon flux; the BL05A1 data shown in Fig. 4 were normalized by the spectral weight of the *dd* excitations, due to noticeable photon flux fluctuation at BL05A1 during measurements

Unless specified, all RIXS spectra presented in the main text were taken at a temperature of 10 K with the photon energy of incident x-rays tuned to the maximum of the absorption curve near the Cu $L_3$-edge (*2p-3d* transition) (Fig. S1a). The scattering geometry is sketched in Supplementary Figure 1. Measurements along the two high symmetry directions $(0,0)$-$(\pi,0)$ and $(0,0)$-$(\pi,\pi)$ were performed on different crystals at the same doping levels and synthesized in the same batch. The crystallographic orientation was aligned using Laue diffraction prior to our RIXS measurements. For the measurements along $(0,0)$-$(\pi,0)$, the (100) direction is aligned in the scattering plane; and for the measurements along $(0, 0) - (\pi, \pi)$, the (110) direction is aligned in the scattering plane. As the electronic structures in NCCO can be regarded as quasi-two-dimensional, the RIXS spectra presented here were plotted as a function of in-plane momentum only ($q_{//}$ shown in Supplementary Figure 1).

We also emphasize that the observed low energy collective excitations do not originate from the ~ 1 % $(Nd,Ce)_2O_3$ secondary phase generated during the annealing process which is required to make the as-grown crystal superconducting[32], because the spectral weight of the modes is significant and Cu $L_3$-edge RIXS is only sensitive to the Cu valence states.

**References:**

1. Kastner, M. A., Birgeneau, R. J., Shirane, G. & Endoh, Y. Magnetic, transport, and optical properties of monolayer copper oxides. *Rev. Mod. Phys.* **70**, 897-928 (1998).

2. Takagi, H. Uchida, S. & Tokura, Y. Superconductivity produced by electron doping in $CuO_2$-layered compounds. *Phys. Rev. Lett.* **62**, 1197-1200 (1989)

3. Scalapino, D. J. A common thread: The pairing interaction for unconventional superconductors. *Rev. Mod. Phys.* **84**, 1383-1417 (2012).

4. Dahm, T. *et al.* Strength of the spin-fluctuation-mediated pairing interaction in a high-temperature superconductor. *Nature Phys.* **5**, 217-221 (2009).

5. Anderson, P. W. Is there glue in cuprate superconductors? *Science* **316**, 1705 (2007).

6. Le Tacon, M. *et al.* Intense paramagnon excitations in a large family of high-temperature superconductor, *Nature Phys.* **7**, 725-730 (2011).

7. Dean, M. P. M. *et al.* Persistence of magnetic excitations in $La_{2-x}Sr_xCuO_4$ from the undoped insulator to the heavily overdoped non-superconducting metal. *Nature Mat.* **12**, 1019-1023 (2013).

8. Armitage, N. P., Fournier, P. & Greene, R. L. Progress and perspectives on electron-doped cuprates. *Rev. Mod. Phys*. **82**, 2421-2487 (2010).

9. Fujita, M. *et al.* Progress in neutron scattering studies of spin excitations in high-$T_C$ cuprates. *J. Phys. Soc. Jpn.* **81**, 011007 (2012).

10. Ament, P. L., van Veenendaal, M., Devereaux, T. P., Hill, J. P. & van den Brink, J., Resonant inelastic x-ray scattering studies of elementary excitations. *Rev. Mod. Phys.* **83**, 705-767 (2011).

11. Strocov, V. N. *et al.* High-resolution soft x-ray beamline ADRESS at the Swiss Light Source for resonant inelastic x-ray scattering and angle-resolved photoelectron spectroscopies. *J. Sync. Rad.* **17**, 631 (2010).

12. Motoyama, E. M. *et al.* Spin correlations in electron-doped high transition-temperature superconductor $Nd_{2-x}Ce_xCuO_4$. *Nature* **445**, 186-189 (2007).

13. Ghiringhelli, G. *et al.* Long range incommensurate charge fluctuations in $(Y, Bd)Ba_2Cu_3O_{6.+x}$. *Science* **337**, 821-825 (2012).

14. Chang, J. *et al.* Direct observation of competition between superconductivity and charge density wave order in $YBa_2Cu_3O_{6.67}$. *Nature Phys.* **8**, 871-876 (2012).

15. Da Silva Neto, E. H. *et al.*, Ubiquitous interplay between charge order and high temperature superconductivity in cuprates. *Science* **343**, 393-396 (2013).

16. Li, Yuan *et al.* Two ising-like magnetic excitations in a single-layer cuprate superconductor. *Nature Phys.* **8**, 404-410 (2012).

17. Braicovich, L. *et al.* Momentum and polarization dependence of single-magnon spectral weight for Cu $L_3$-edge resonant inelastic x-ray scattering from layered cuprates. *Phys*. Rev. B **81**, 174533 (2010).

18. Guarise, M. *et al.* Measurement of magnetic excitations in the two-Dimensional antiferromagnetic $Sr_2CuO_2Cl_2$ insulator using resonant x-Ray scattering: evidence for extended interactions. *Phys. Rev. Lett.* **105**, 157006 (2010).

19. Bourges, P., Casalta, H., Ivanov, A. S. & Petitgrand, D. Superexchange coupling and spin susceptibility spectral weight in undoped monolayer cuprates. *Phys. Rev. Lett.* **79**, 4906 (1997).

20. Wilson, S. D. *et al.* High-energy spin excitations in the electron-doped superconductor $Pr_{0.88}LaCe_{0.12}CuO_{4-\delta}$ with $T_c = 21$ K. *Phys. Rev. Lett.* **96**, 157001 (2006).

21. Fujita, M., Matsuda, M., Fåk, B, Frost, C. D. & Yamada, K. Novel spin excitations in optimally electron-doped $Pr_{0.89}LaCe_{0.11}CuO_4$. *J. Phys. Soc. Jpn.* **75**, 093704 (2006).

22. F. Schmitt *et al.* Analysis of the spectral function of $Nd_{1.85}Ce_{0.15}CuO_4$ obtained by angle-resolved photoemission spectroscopy. *Phys. Rev. B* **78**, 100505 (2008).

23. Singley, E. J., Basov, D. N., Kurahashi, K., Uefuji, T. & Yamada. K. Electron dynamics in $Nd_{1.85}Ce_{0.15}CuO_{4+\delta}$. *Phys. Rev. B* **64**, 224503 (2001).

24. Yu, G. *et al.,* Two characteristic energies in the low-energy magnetic response of the electron-doped high-temperature superconductor $Nd_{2-x}Ce_xCuO_{4+\delta}$. *Phys. Rev. B* **82**, 172505 (2010).

25. Matsui, H. *et al.* Evolution of the pseudogap across the magnet-superconductor phase boundary of $Nd_{2-x}Ce_xCuO_4$. *Phys. Rev. B* **75**, 224514 (2007).

26. Hinton, J. P. *et al.* Time-resolved optical reflectivity of the electron-doped $Nd_{2-x}Ce_xCuO_{4+\delta}$ cuprate superconductor: evidence for an interplay between competing orders. *Phys. Rev. Lett.* **110**, 217002 (2013).

27. Chakravarty, S., Laughlin, R. B., Morr. D. K. & Nayak, C. Hidden order in the cuprates. *Phys. Rev. B* **63**, 094503 (2001).

28. Varma, C. M. Theory of pseudogap state of the cuprates. *Phys. Rev. B* **73**, 155113 (2006).

29. Broun, D. M., What lies beneath the dome? *Nature Phys.* **4**, 170-172 (2008).

30. Vishik, I. M. *et al.* Phase competition in trisected superconducting dome. *PNAS* **109**, 18332-18337 (2012).

**Supplementary Information** is enclosed with this submission.

**Acknowledgements:** The authors appreciate Yang Lu's support for characterizing the doping levels of the measured samples. This work was supported by the U. S. Department of Energy, Office of Basic Energy Science, Division of Materials Science and Engineering under the contract no. DE-AC02-76SF00515. The work at University of Minnesota was supported by the NSF and the NSF MRSEC program. S.G. acknowledges support from the Swiss NSF (Contract No. P2EZP2_148737). The authors appreciate the experimental supports from the ADRESS beam line of the Swiss Light Source (SLS) at the Paul Scherrer Institut, Switzerland and the beam line BL05A1 at the National Synchrotron Radiation Research Center (NSRRC), Taiwan.

**Author Contributions:** W.S.L conceived and designed the experiments with suggestions from Z.X., M.G., T.P.D. and T.S. W.S.L, J.J.L., W.T., S.G., S.W.H., Y.B.H., V.N.S., and T.S. performed the measurement at the SLS. W.S.L., H.Y.H., R.P.W., W.B.W. and D.J.H. performed the measurement at the NSRRC. E.M.M., G.Y., and M.G. synthesized and prepared the single crystals used for the measurements. E.A.N., B.M., and T.P.D. performed the theoretic calculations. W.S.L. wrote the manuscript with contributions from all authors.

**Competing Financial Interest:** The authors declare no competing financial interests.

**Correspondence** and requests for materials should be addressed to leews@stanford.edu, zxshen@stanford.edu and tpd@stanford.edu

**Figure 1 Magnons in antiferromagnetic NCCO ($x$ = 0.04). a,** Energy-momentum intensity maps (upper) and waterfall plots (lower) of representative RIXS spectra along (0, 0) - ($\pi$, 0) and (0, 0) - ($\pi$, $\pi$), respectively. Here momenta are expressed in units of the reciprocal lattice spacing (1/$a$). Red ticks indicate the maxima of the spectra, which are dominated by single magnon excitations. The data were taken with $\sigma$ polarization light. **b,** Representative spectra taken with $\pi$ and $\sigma$ incident x-ray polarizations at symmetric $q_{//}$ positions Thick black curves are from the sum of Gaussian fits to the elastic scattering peak as well as phonon (two thin black curves), a single magnon peak (red shaded region) and a constant background (black dashed line). The residual spectral weight after subtracting this fit is shown by the blue curves. The fitted magnon peak positions have been superimposed on the color maps in **a** (red markers) and further summarized in **c**. **c,** Magnon dispersion along (0, 0)–(|$\pi$|, 0) and (0, 0)–($\pi$, $\pi$) deduced from RIXS spectra. Error bars are estimated by the uncertainty in determining the energy loss reference point. The dashed curves show a linear-spin-wave theory dispersion for a two dimensional antiferromagnetic Heisenberg model fitted to the data with $c_s$ = 830 ± 9 meV·Å.

**Figure 2 Paramagnons and unexpected collective modes in superconducting NCCO ($x$ = 0.147). a,** Energy-momentum intensity maps (upper) and waterfall plots (lower) of RIXS spectra along (0, 0)-($\pi$, 0) and (0, 0)-($\pi$, $\pi$), respectively. Red (blue) ticks indicate the peak positions for paramagnons (unexpected collective modes). The data were taken with $\sigma$ polarization light. **b,** Spectra near the Γ point with finer momentum steps highlight the dispersion of the unexpected collective modes. **c,** RIXS spectra taken with $\pi$ and $\sigma$ incident x-ray polarization at symmetric $q_{//}$

points. The red and blue shaded areas are Gaussian fits to the paramagnons and the collective modes, respectively. The thick black curve is a fit to the data. The thin black and dashed black curves are a Gaussian fit to the elastic scattering signal and the background as determined from the *dd* and charge-transfer excitations at higher energy loss. **d,** Paramagnon and collective mode dispersions along $(0, 0) – (|\pi|, 0)$ and $(0, 0) - (\pi, \pi)$ directions deduced from RIXS spectra. The paramagnon dispersion deviates from simple linear-spin-wave theory (as a guide-to-eye, black dashed curve, with $c_s = 1240$ meV). For comparison, the magnon dispersion in the $x = 0.04$ compound is also superimposed. The blue dashed line is a fit to the collective mode dispersion in the form of generic collective charge excitations as described in the text.

**Figure 3 Magnetic excitations hardening. a,** RIXS spectra of $x = 0.04$ and $0.147$ compounds at two representative in-plane momentum points as indicated by the red markers in the insets. Red ticks indicate the positions of the paramagnons and the collective modes, respectively. **b,** Quantum Monte Carlo simulation of the dynamic spin structure factor $S(q, \omega)$ at $(\pi/2, \pi/2)$ in the single band Hubbard model for both electron (upper) and hole-doping (lower). Red ticks indicate the positions of maximum intensity of $S(q, \omega)$.

**Figure 4 Doping and temperature dependence of the collective modes. a,** RIXS spectra of the $x = 0.166$ compound along the $(0, 0) - (\pi, 0)$ direction. Red (blue) ticks indicate the peak positions for paramagnons (unexpected collective modes). The data were taken with σ polarization light. **b,** Energy-momentum dispersion of the collective modes in superconducting compounds $x = 0.147$ (open symbol) and $x = 0.166$ (close symbol). The shaded area indicated the energy-momentum region that was not resolvable due to finite resolution of the instrument. **c,** Temperature dependent RIXS spectra taken at a momentum position near the Γ point for $x =$

0.147 (upper) and $x = 0.166$ (lower) compounds. **d,** The temperature dependent spectral weight of the collective modes, shown in **c,** is calculated by integrating the background-subtracted spectra between 0.2 and 0.4 eV. As motivated by the $T = 270$ K spectra of the $x = 0.166$ compound, the background is assumed to be energy independent with the value of the spectral intensity at 1 eV. The plotted temperature dependent spectral weight is normalized to that of the lowest temperature. The error bars are estimated by the noise level of the spectra. The temperature dependent spectral weight is also plotted in the phase diagram (right panel) via color intensity stripes to demonstrate such drastic change of temperature dependence within a small doping range. The color scale to the normalized spectral weight is indicated as the color scale bar.

**Figure 5 Distinct doping evolution of the collective excitations in electron- and hole-doped cuprates.** Sketches of collective excitation spectra in the energy-momentum space for electron-doped (e-SC, NCCO), lightly doped antiferromagnetic (AFM), and hole-doped (h-SC) superconducting cuprates. Representative data points shown in Figs.1 and 2 have been superimposed on the panels for e-SC. The bandwidth of the magnetic excitations is marked in the energy axis. The low energy magnetic excitations along $(\pi,\pi)$-$(0, \pi)$ direction near $(\pi, \pi)$ obtained via inelastic neutron scattering are sketched in the images. The spin gap ($\Delta_S$) and spin incommensurability ($\varepsilon$) are not drawn to scale. The lowest panel is a sketch of the cuprate phase diagram. The dots represent the quantum critical points associated with symmetry-broken states on both sides of the phase diagram. While the existence of such quantum critical point in electron-doped side of phase diagram is suggested by our results, its exact position in doping remains an open question.

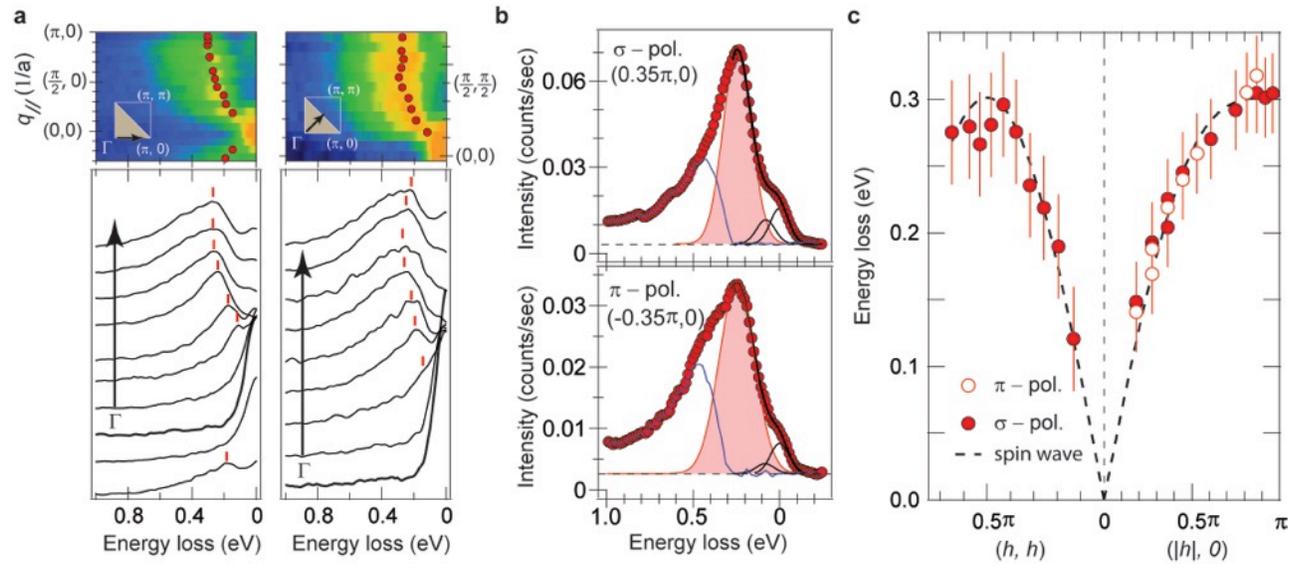

Figure 1

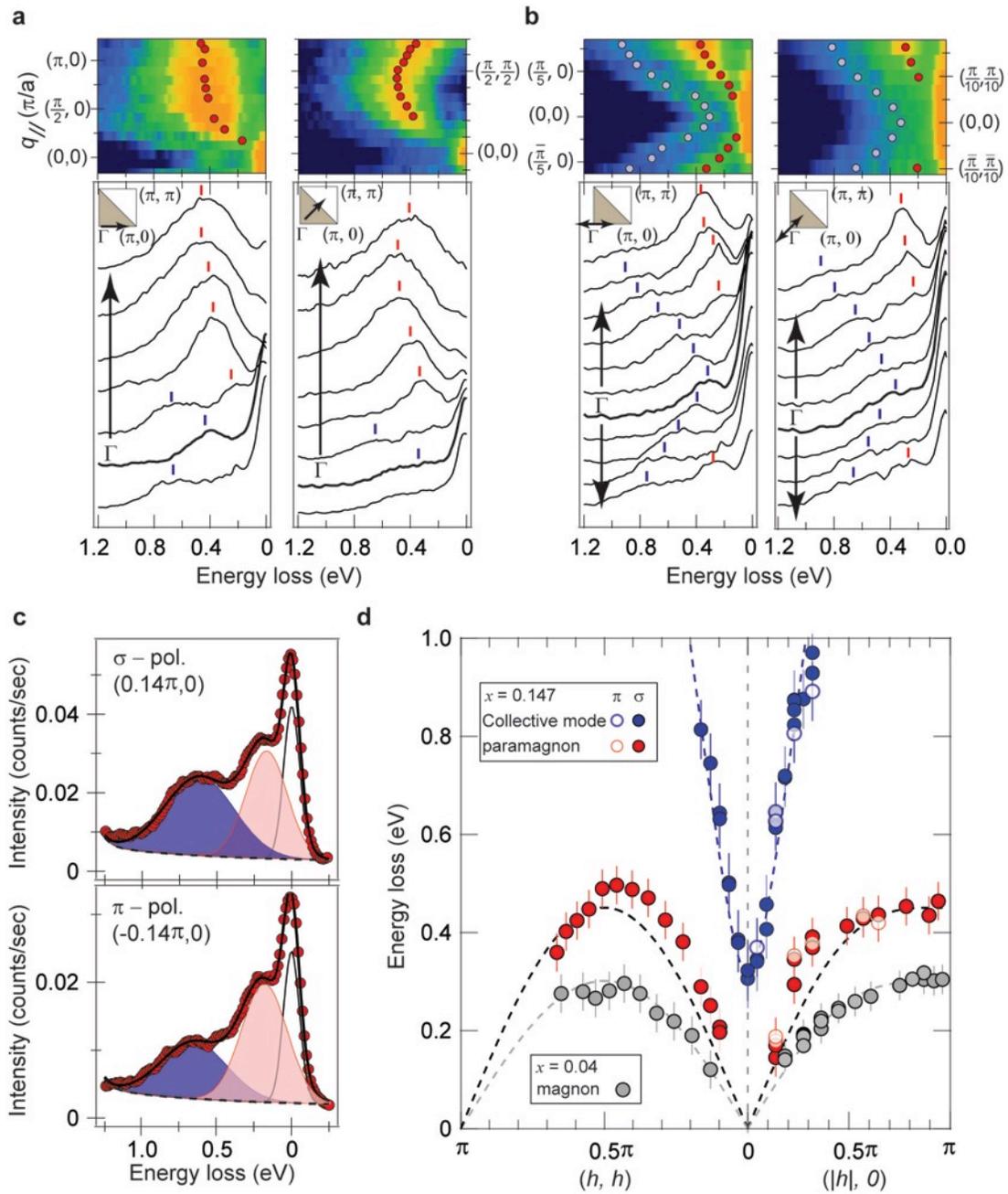

Figure 2

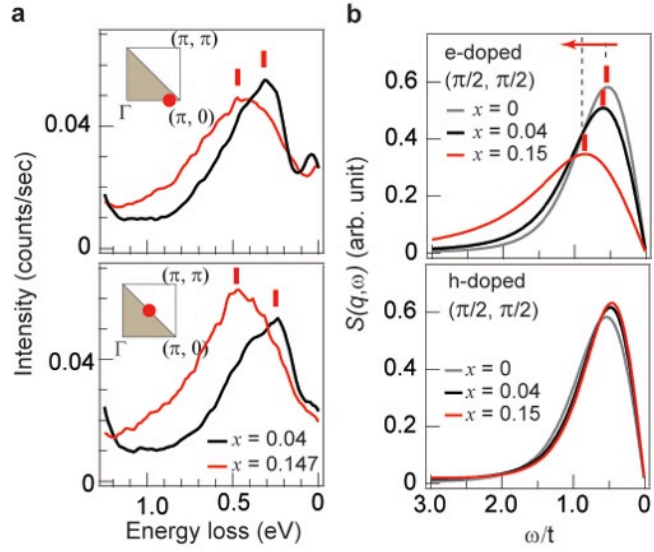

Figure 3

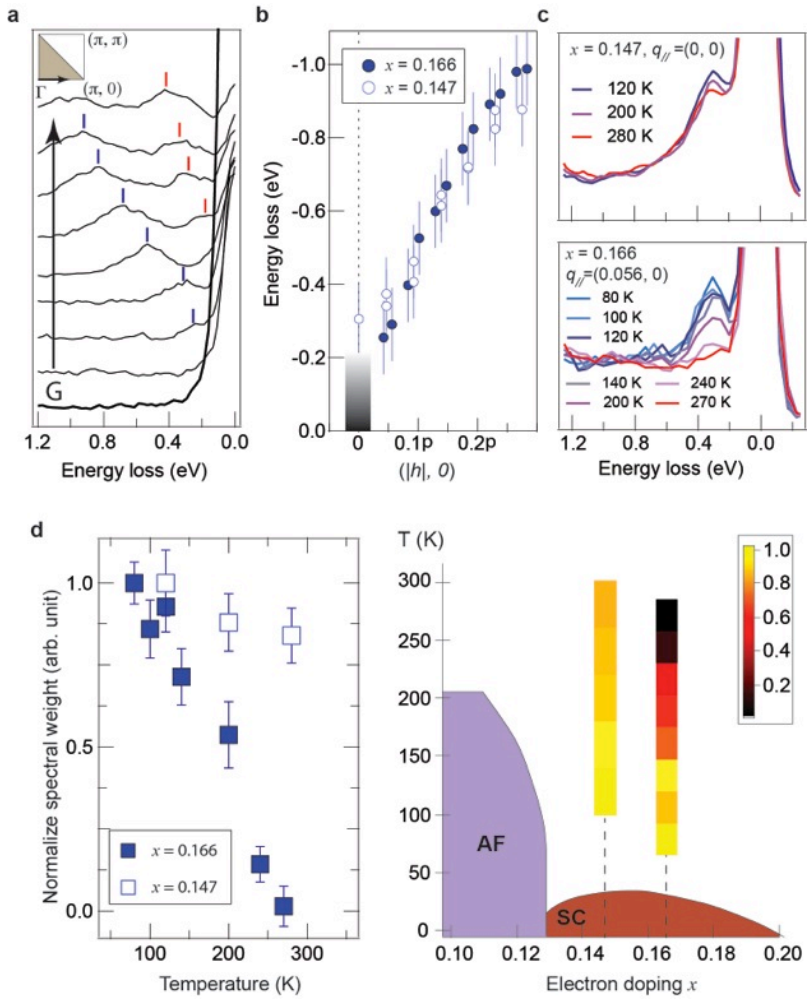

Figure 4

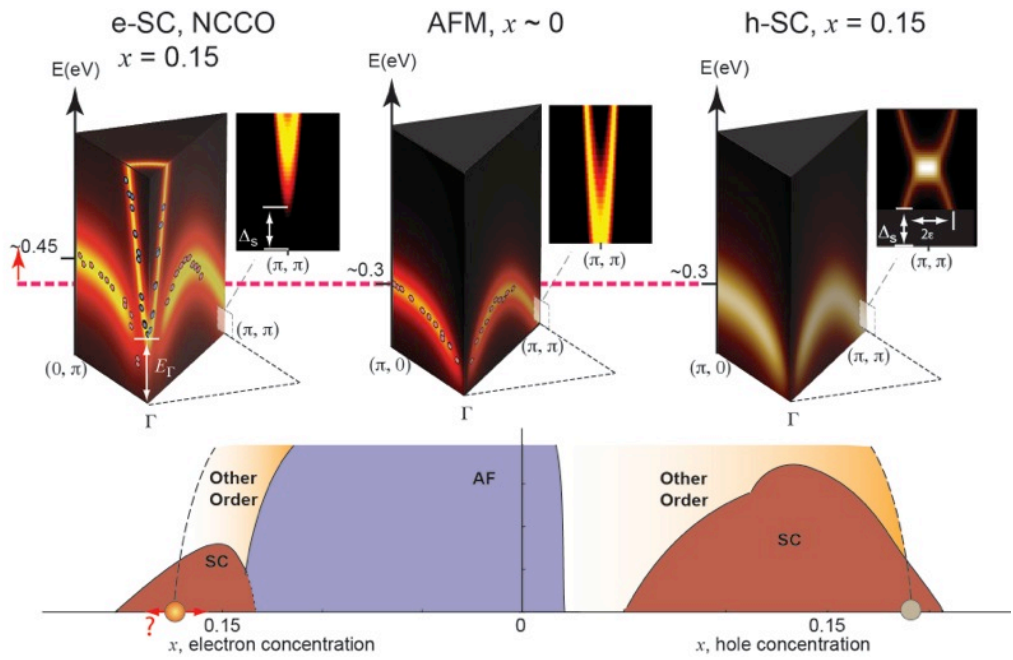

Figure 5